\documentclass[acmtocl]{acmtrans2m}
\title{Weyl's Predicative Classical Mathematics as a Logic-Enriched Type Theory}
\author{ROBIN ADAMS and ZHAOHUI LUO \\ Dept of Computer Science, Royal Holloway, Univ of London}
\newcommand{\LTTW}{\ensuremath{\mathrm{LTT}_\mathrm{w}}}

\begin{abstract}
We construct a logic-enriched type theory \LTTW~that corresponds closely to the predicative system of foundations presented by Hermann Weyl in
\emph{Das Kontinuum}.
We formalise many results from that book in \LTTW,
including Weyl's definition of the cardinality of a set and
several results from real analysis, using
the proof assistant Plastic that implements the logical framework LF.
This case study shows how type theory can be used to represent a
non-constructive foundation for mathematics.
\end{abstract}


\newcommand{\vald}{\ \mathrm{valid}}

\newcommand{\El}[1]{El \left( {#1} \right)}
\newcommand{\Type}{\mathbf{Type}}
\newcommand{\Prop}{\mathbf{Prop}}
\newcommand{\Prf}[1]{Prf \left( {#1} \right)}
\newcommand{\kind}{\ \mathrm{kind}}

\newcommand{\Set}[1]{\mathrm{Set} \left( {#1} \right)}
\newcommand{\s}{\mathrm{s}}
\newcommand{\EN}{\mathrm{E}_\mathbb{N}}

\newcommand{\IN}{\mathrm{Ind}_\mathbb{N}}
\newcommand{\IX}{\mathrm{Ind}_\times}
\newcommand{\IA}{\mathrm{Ind}_{\boldarrow}}
\newcommand{\LF}{LF$'$}
\newcommand{\p}{\mathrm{prop}}
\newcommand{\set}{\mathrm{set}}
\newcommand{\boldarrow}{\Rightarrow}

\newcommand{\selfcomment}[1]{\ifodd 0 {\sf #1 }\fi}

\usepackage{float}
\usepackage{amsbsy}
\usepackage{amsmath}
\usepackage{amssymb}
\usepackage{diagrams}
\usepackage{prooftree}

\newarrow{Line}-----
\newarrow{Etc}.....

\begin{document}
\begin{bottomstuff}
Authors' email addresses: \texttt{\{robin,zhaohui\}@cs.rhul.ac.uk}

This work was supported by the UK EPSRC research grants
GR/R84092, GR/R72259, EP/D066638/1, the UK Leverhulme grant F/07-537/AA, and EU TYPES grant 510996.
\end{bottomstuff}

\maketitle
\floatstyle{boxed}
\restylefloat{figure}

\section{Introduction}
Since the start of the 20th century, many different formal logical systems have been proposed as foundations for mathematics.  Each of these foundational systems provides a formal language in which the statements of mathematics can be written, together with axioms and rules of deduction by which, it is claimed, the theorems of mathematics can be deduced.  These different systems were often motivated by different philosophical schools of thought on the question of which proofs and constructions in mathematics are valid.  These schools include \emph{constructivism}, which holds that the use of the axiom of excluded middle is illegitimate, and \emph{predicativism}, which holds that definitions that involve a certain form of circularity are illegitimate.

\emph{Proof assistants} or \emph{proof checkers} are tools that help the user to construct formal proofs of theorems in these formal logical systems, and provide a guarantee of each proof's correctness.  The logical systems known as \emph{type theories} have proven particularly successful for this purpose.  However, so far,
the type theories that have been used have almost exclusively been \emph{constructive}.  Examples include the proof checker Coq \cite{Coq:manual} based on the type theory CIC \cite{coq'art}, and the proof checker Agda \cite{norell:thesis} based on Martin-L\"of Type Theory \cite{pmltt}.
%

We would like to investigate how the other schools in the foundations of mathematics may be formalised with type theories.  As a first case study, we chose the classical, predicative system presented in Hermann Weyl's 1918 text \emph{Das Kontinuum} \cite{weyl:kontinuum}.  In this book, Weyl investigates how much of the mathematical corpus can be retained if we restrict ourselves to predicative definitions and methods of proof.  He presents a foundational system in which it is impossible to perform an impredicative definition.  He proceeds to construct the real numbers and prove many theorems of mathematics within this system.  It is an excellent example of a fully developed non-mainstream foundational system for mathematics.

In the system Weyl presents in \emph{Das Kontinuum}, only \emph{arithmetic} sets may be constructed.  We can form the set
\[ \{ x \mid \phi[x] \} \]
only if the proposition $\phi[x]$ is \emph{arithmetic} --- that is, it does not involve quantification over sets.
Weyl's aim was to investigate how much mathematics we can reconstruct while restricting ourselves to the arithmetic sets.
He shows that many results that are usually proven impredicatively can be proven predicatively; and
that, even for those results that cannot, one can often prove a weaker result which in practice is just as useful.

We have constructed a system $\LTTW$ that corresponds very closely to the semi-formal system that Weyl presented.  Our system $\LTTW$ is a \emph{logic-enriched type theory} (LTT), a type theory extended with a separate mechanism for forming and proving propositions.  It contains types of natural numbers, ordered pairs and functions; classical predicate logic with equality, together with the ability to prove propositions by induction; and the ability to form sets by predicative definition.

We have used the proof assistant Plastic to formalise in $\LTTW$ many of the theorems and proofs presented in Weyl~\shortcite{weyl:kontinuum}.  The work presented here forms a case study in how type theory --- specifically logic-enriched type theories --- may be used outside the realm of constructive mathematics, to construct foundational systems in such a way that it is practicable to carry out machine-supported formalisation of proofs.

\subsection{Outline}

In Section \ref{section:background}, we give some background on logic-enriched type theories and the historical context to Weyl's work.
In Section \ref{section:weyl}, we describe in
detail the version of Weyl's foundational system we shall be using.
We proceed in Section \ref{section:lett} to describe a
logic-enriched type theory within a modified version of the logical
framework LF\footnote{The logical framework LF here is a Church-typed
version of Martin-L\"{o}f's logical framework \cite{pmltt}, and
is not to be confused with the Edinburgh LF \cite{ELF:93}.  Among other
differences, LF allows the user to declare computation rules,
and hence to specify type theories such as Martin-L\"{o}f's type theory
\cite{pmltt} and UTT \cite{luo:car}.} \cite{luo:car}.  We
claim that this logic-enriched type theory faithfully corresponds to
the system presented in the preceding section.
In Section \ref{section:formal}, we describe the results proven in the formalisation, which was carried out
in a modified version of the
proof assistant Plastic \cite{plastic}, an implementation of LF.
In Section \ref{section:predicativity}, we describe some of the other approaches to predicativity that have been followed, and discuss how they might be formalised in a similar way.

A preliminary version of this paper appeared in the proceedings of TYPES 2006 \cite{al:wpcm}.
The source code of the formalisation, together with a list of all the definitions and results in \emph{Das Kontinuum}, is
available at \\
\verb;http://www.cs.rhul.ac.uk/~robin/weyl;

\section{Background}
\label{section:background}
\subsection{Type Theories for Non-constructive Mathematics}

\begin{figure}
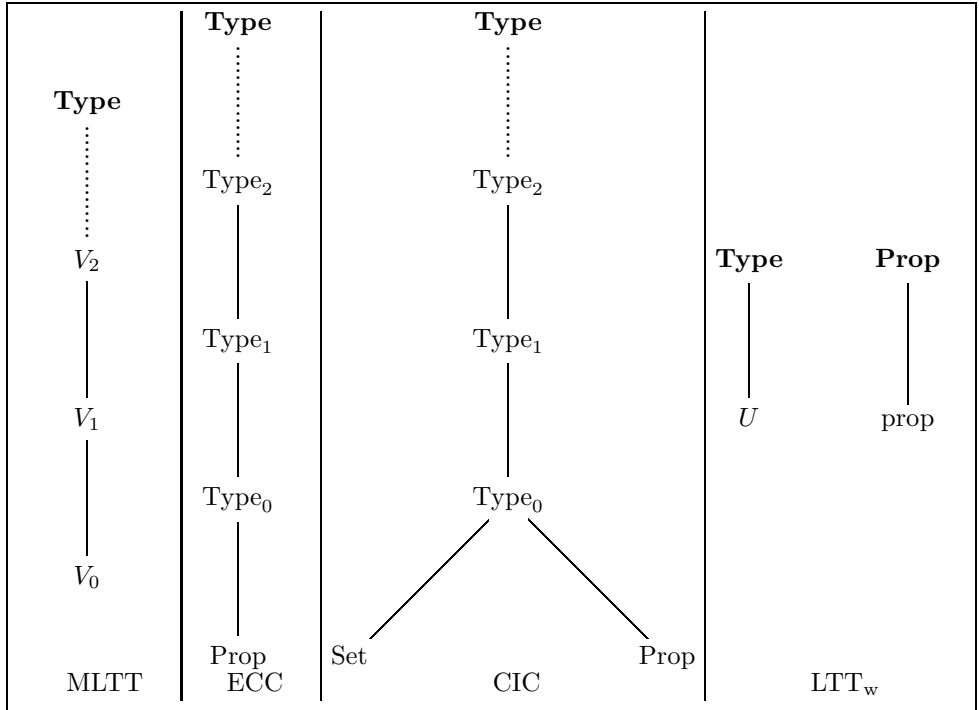

\begin{tabular}{*3{c@{\hspace{.2in}}|@{\hspace{.05in}}}c}
\begin{diagram}
\Type \\
       \uEtc \\
V_2 \\
\uLine \\
V_1 \\
\uLine \\
V_0
      \end{diagram} &

\begin{diagram}
\Type \\
        \uEtc \\
\mathrm{Type}_2 \\
\uLine \\
\mathrm{Type}_1 \\
\uLine \\
\mathrm{Type}_0 \\
\uLine \\
\mathrm{Prop}
       \end{diagram}
& \begin{diagram}
& & \Type & & \\
& &          \uEtc & & \\
& & \mathrm{Type}_2 & & \\
& & \uLine & & \\
& & \mathrm{Type}_1 & & \\
& & \uLine & &\\
& & \mathrm{Type}_0 & & \\
& \ruLine & & \luLine & \\
\mathrm{Set} & & & & \mathrm{Prop}
        \end{diagram}
& \begin{diagram}
\Type & & \Prop \\
\uLine & & \uLine \\
U & & \p
        \end{diagram} \\
~~~~MLTT & ~~~~ECC & ~~~~CIC & ~~~~\LTTW
\end{tabular}
\caption{The division of types into propositions and datatypes in several different type theories.  A universe is placed under another, joined by a line, if the first is an object of the second.  The kind $\Type$ is placed above the type universes, and the kind $\Prop$ above the propositional universe of $\LTTW$.}
\label{fig:Fig1}
\end{figure}

A type theory divides mathematical objects into \emph{types}.  The types themselves are often collected into \emph{universes}.
The usual method for using a type theory as a logical system is known as \emph{propositions as types}: some or all of the types are identified with propositions, and the objects of each type with proofs of that proposition.  We prove a theorem by constructing an object of the appropriate type.  In Martin-L\"of Type Theory (MLTT) \cite{pmltt}, every type is considered a proposition.  In other type theories, such as ECC \cite{luo:car} or CIC, the basis for the proof checker Coq \cite{Coq:manual}, only some of the types are considered propositions, and these are collected into a universe, usually denoted by $\mathrm{Prop}$.  The other types are often called \emph{datatypes} to distinguish them.  Figure \ref{fig:Fig1} shows the universe structure of several type theories.

When types are identified with propositions in this way, many natural type constructions correspond to the connectives of intuitionistic logic.  For example, the non-dependent product $A \times B$ corresponds to conjunction ($A \wedge B$), and the dependent product $\Pi x:A.B$ corresponds to universal quantification ($\forall x:A.B$).  That is, the introduction and elimination rules for $A \times B$ mirror the introduction and elimination rules for conjunction in intuitionistic logic; similarly, the rules for dependent product mirror those for universal quantification.
Type theories are thus very well suited for formalising intuitionistic mathematics.

There are several ways in which a type theory may be modified so as to be appropriate for formalising classical mathematics. This cannot however be done without changing the structure of the datatypes, because the two interact so strongly.  In MLTT, they are one and the same; in ECC or CIC, the universes $\mathrm{Type}_0$, $\mathrm{Type}_1$,\nolinebreak \ldots contain both propositions and datatypes.

It is possible to introduce constructions
into the type theory so that the theory's rules now mirror the rules of deduction of classical logic, such as the `freeze' and `unfreeze' operations of the $\lambda \mu$-calculus \cite{lambdamu}.  However, doing so allows new objects to be formed in the datatypes.

There have also been several formalisations of classical proofs which used an intuitionistic type theory with additional axioms,
such as Gonthier's proof of the Four Colour Theorem
\cite{gonthier:fct}, which extended the proof checker Coq with axioms for the real numbers that imply the axiom of excluded middle.  However, this approach introduces non-canonical objects into the datatypes.
(Further discussion on these points can be found in \cite{luo:LTT06}.)



This problem does not arise in the systems known as \emph{logic-enriched type theories} (LTTs) introduced by Aczel and Gambino \cite{ag:cpdtt,ga:gticst}.  These are type theories in which the propositions and the datatypes are completely separate.  It is thus possible to introduce axioms or new rules of deduction without affecting the datatypes.  We shall construct such an LTT in this paper, which we shall call \LTTW.  Its universe structure is also shown in Figure \ref{fig:Fig1}.

There are several features of type theory that are of especial benefit for proof assistants: each object carries
a type which gives information about that object, and the type theory itself has a primitive notion of computation.
We contend that the intuitions behind type theory apply outside of
intuitionistic mathematics, and that these advantages
would prove beneficial when applied to other forms of proof.  It is
equally natural in classical mathematics to divide mathematical
objects into types, and it would be of as much benefit to take
advantage of the information provided by an object's type in a
classical proof.  The notion of computation is an important part of
classical mathematics.  When formally proving a property of a
program, we may be perfectly satisfied with a classical proof, which
could well be shorter or easier to find.

We further contend that it is worth developing and studying type
theories specifically designed for non-constructive mathematical
foundations.  For this purpose,
logic-enriched type theories would seem to be particularly appropriate.

\subsection{Logic-Enriched Type Theories}

The concept of an LTT, an extension of the notion of type theory,
was proposed by Aczel and Gambino in their study of type-theoretic
interpretations of constructive set theory
\cite{ag:cpdtt,ga:gticst}.
A \emph{type-theoretic framework}, which
formulates LTTs in a logical framework, has been proposed in
\cite{luo:LTT06} to support formal reasoning with different logical
foundations.  In particular, it adequately supports classical
inference with a predicative notion of set, as described below.


An LTT consists of a type theory augmented with a separate,
primitive mechanism for forming and proving propositions.  We
introduce a new syntactic class of \emph{formulas}, and new
judgement forms for a formula being a well-formed proposition, and
for a proposition being provable from given hypotheses.

An LTT thus has two rigidly separated `worlds': the
\emph{datatype} world of terms and types, and the \emph{logical}
world of proofs and propositions, for describing and
reasoning about the datatype world.  This provides two advantages over traditional type theories:

\begin{longitem}
\item
We have separated the datatypes from the propositions.  This allows us to add axioms without changing the datatype world.  We can, for example, add the axiom of excluded middle without thereby causing all the datatypes of the form $A + (A \rightarrow \emptyset)$ to be inhabited.
\item
We do not have any computation rules on proofs.  Further, a proof cannot occur inside a term, type or proposition.  We are thus free to add any axioms we like to the logic: we know that, by adding the axiom of excluded middle (say),  we shall not affect any of the properties of the reduction relation, such as decidability of convertibility or strong normalisation.
\end{longitem}

\subsubsection{Remark} The clear separation between logical
propositions and data types is an important salient feature of LTTs
\cite{ag:cpdtt,ga:gticst} and the associated logical framework
\cite{luo:LTT06}. In Martin-L\"{o}f's type theory, for example,
types and propositions are identified. The second author has argued,
in the development of ECC/UTT \cite{luo:car} as implemented in
Lego/Plastic \cite{legomanual:Report,plastic}, that it is unnatural
to identify logical propositions with data types and there should be
a clear distinction between the two. This philosophical idea was
behind the development of ECC/UTT, where propositions are types, but not all types are propositions.
LTTs have
gone one step further -- propositions and types are separate syntactic categories.

\subsection{Foundations of Mathematics}

When building a foundational system for mathematics, two of the decisions that must be made are:
\begin{longenum}
 \item Whether the logic shall be \emph{classical} or \emph{intuitionistic}.  In intuitionistic logic, principles such as excluded middle ($\phi \vee \neg \phi$) or $\neg \forall x \phi(x) \rightarrow \exists x \neg \phi(x)$ are not universally valid.  A proof of $\phi \vee \psi$ must provide a way of deciding which of $\phi$ or $\psi$ holds, and a proof of $\exists x \phi(x)$ must provide a way of constructing an $x$ for which $\phi(x)$ holds.
\item Whether \emph{impredicative} definitions are allowed, or only \emph{predicative}.  A definition is \emph{impredicative} if it involves a certain kind of `vicious circle', in which an object is defined in terms of a collection of which it is a member.  (A detailed discussion of the concept of predicativity and its history is given in Section \ref{section:predicativity}.)
\end{longenum}

Each of the four possible combinations of these
options has been advocated as a foundation for mathematics at
some point in history.
\begin{longitem}
\item
\textbf{Impredicative classical mathematics}. This is arguably the way in
which the vast majority of practising mathematicians work.
Zermelo-Fraenkel Set Theory (ZF) is one such foundation.  The proof checker Mizar \cite{muzalewski:mizar} has been used to formalise a very large body of impredicative classical mathematics.
\item
\textbf{Impredicative constructive mathematics}.  Impredicative
type theories such as CC \cite{ch:coc}, UTT \cite{luo:car}, and CIC \cite{coq'art}
are examples of such foundations.  These have been implemented by the proof checkers LEGO \cite{legoWWW} and Coq \cite{Coq:manual}.
\item
\textbf{Predicative classical mathematics}.  This was the approach
taken by Weyl in his influential monograph of 1918, \emph{Das
Kontinuum} \cite{weyl:kontinuum}.  Stronger predicative classical systems have been investigated by Feferman \cite{feferman:spa} and Sch\"utte \cite{schutte:pwo}.
\item
\textbf{Predicative constructive mathematics}.  Its foundations are
provided, for example, by Martin-L\"{o}f's type theory
\cite{pmltt,ml:itt}.
\end{longitem}

LTTs may provide a uniform type-theoretic framework that can
support formal reasoning with these four different logical foundations and others.  This idea is discussed further in
\cite{luo:LTT06}.

In this paper, we present a case study in the type-theoretic
framework: to construct an LTT to represent the predicative, classical foundational system of mathematics
developed by Weyl in his monograph \textit{Das Kontinuum}
\cite{weyl:kontinuum}, and to formalise in that LTT several of
the results proven in the book.

The system presented in the
book has since attracted interest, inspiring for example the
second-order system ACA$_0$ \cite{feferman:kontinuum}, which plays
an important role in the project of Reverse Mathematics
\cite{simpson:sosoa}.  It is a prominent example of a fully
developed non-mainstream mathematical foundation, and so a
formalisation should be of quite some interest.

\section{Weyl's Predicative Foundations for Mathematics}
\label{section:weyl}

Hermann Weyl (1885--1955) contributed to
many branches of mathematics in his lifetime.  His greatest
contribution to the foundations of mathematics was the book
\emph{Das Kontinuum} \cite{weyl:kontinuum} in 1918, in which he
presented a predicative foundation which he showed was
adequate for a large body of mathematics.

The semi-formal presentation of the foundational system in \emph{Das Kontinuum} would not be acceptable by modern standards.  Weyl does not give a rigorous formal definition of his syntax, axioms or rules of deduction.  In this section, we shall give a formal definiton of a modern reconstruction of Weyl's foundational system.

\pagebreak

The notation of our system shall differ considerably from Weyl's own.
We shall also include several features not present in Weyl's system which are redundant in theory, but very convenient practically.
The differences between our system and Weyl's shall be discussed under Section \ref{section:extension}
below.

\subsection{Weyl's Foundational System}

Weyl's system is constructed according to these principles:
\begin{enumerate}
\item
The natural numbers are accepted as a primitive concept.
\item
Sets and relations can be introduced by two methods: explicit
definitions, which must be \emph{predicative}; and definition by
recursion over the natural numbers.
\item
Statements about these objects have a definite truth value; they are either true or false.
\end{enumerate}

Regarding point 2, we are going to provide ourselves with the ability to define sets by \emph{abstraction}: given a formula $\phi[x]$ of the system, to form the set
\begin{equation}
 \label{abs}
 S = \{ x \mid \phi[x] \} \enspace .
\end{equation}
In order to ensure that every such definition is predicative, we restrict which quantifiers can occur in the formula $\phi[x]$ that can appear in (\ref{abs}): we may quantify over natural numbers, but we may not quantify over sets or functions.  In modern terminology, we would say that $\phi[x]$ must be \emph{arithmetic}; that is, it must contain only first-order quantifiers.

\subsubsection{Components of Weyl's System}

Weyl divides the universe of mathematical objects into collections which he calls \emph{categories}.  The categories are divided into \emph{basic} categories and \emph{ideal} categories.  Each category has \emph{objects}.
There are also \emph{propositions}, which are divided into the
\emph{arithmetic}\footnote{Weyl chose the German word \emph{finite}, which in other contexts is usually translated as `finite'; however, we agree with Pollard and Bole \cite{weyl:continuum} that this would be misleading.} propositions, and the \emph{large} propositions.

\subsubsection{Categories}
\begin{enumerate}
\item There is a basic category $\mathbb{N}$, whose objects are called \emph{natural numbers}.
\item For any two categories $A$ and $B$, there is a category $A \times B$, whose objects are called \emph{pairs}.  If $A$ and $B$ are basic categories, then $A \times B$ is a basic category; otherwise, $A \times B$ is ideal.
\item For any two categories $A$ and $B$, there is a category $A \boldarrow B$, whose objects are called \emph{functions}.  The category $A \boldarrow B$ is always an ideal category.
\item For any category $A$, there is a category $\Set{A}$, whose objects are called \emph{sets}.  The category $\Set{A}$ is always an ideal category.
\end{enumerate}
For applications to other branches of mathematics, the system may be extended with other basic categories.  For example, when formalising geometry, we may include a basic category of points and a basic category of lines.

\subsubsection{Objects}
For each category $A$, there is a collection of \emph{variables associated with} $A$.
\begin{enumerate}
 \item Every variable associated with a category $A$ is an object of category $A$.
\item There is an object 0 of the category $\mathbb{N}$.
\item For every object $n$ of the category $\mathbb{N}$, there is an object $\s(n)$ of the category $\mathbb{N}$, the \emph{successor} of $n$.
\item For every object $a$ of category $A$ and $b$ of category $B$, there is an object $(a,b)$ of category $A \times B$.
\item For every object $p$ of category $A \times B$, there is an object $\pi_1(p)$ of category $A$ and an object $\pi_2(b)$ of category $B$.
\item Let $x$ be a variable associated with category $A$, and $b$ an object of category $B$.  Then there is an object $\lambda x.b$ of category $A \boldarrow B$.
\item For every object $f$ of category $A \boldarrow B$ and object $a$ of category $A$, there is an object $f(a)$ of category $B$.
\item Let $x$ be a variable associated with category $A$, and $\phi$ an \emph{arithmetic} proposition.  Then there is an object $\{ x \mid \phi \}$ of category $\Set{A}$.
\item Let $f$ be an object of category $A \boldarrow A$.  Then there is an object $\overline{f} : A \times \mathbb{N} \boldarrow A$, the \emph{iteration} of $f$.  (Intuitively, $\overline{f}((x,n))$ is the result of applying $f$  to $x$, $n$ times.)
\end{enumerate}

\subsubsection{Propositions}
\begin{enumerate}
\item
If $A$ is a \emph{basic} category, and $a$ and $b$ are objects of $A$, then there is an arithmetic proposition $a = b$.
\item
If $a$ is an object of category $A$, and $S$ an object of category $\Set{A}$, then there is an arithmetic proposition $a \in S$.
\item
If $\phi$ is a proposition, then $\neg \phi$ is a proposition.  If $\phi$ is arithmetic, then $\neg \phi$ is arithmetic; if $\phi$ is large, then $\neg \phi$ is large.
\item
If $\phi$ and $\psi$ are propositions, then $\phi \wedge \psi$, $\phi \vee \psi$ and $\phi \supset \psi$ are propositions.  If $\phi$ and $\psi$ are arithmetic, then these three propositions are arithmetic; if either $\phi$ or $\psi$ is large, then these three propositions are large.
\item
If $x$ is a variable associated with $A$ and $\phi$ is a proposition, then $\forall x \phi$ and $\exists x \phi$ are propositions.  If $A$ is basic and $\phi$ is arithmetic, then these two propositions are arithmetic.  If $A$ is ideal or $\phi$ is large, then these two propositions are large.
\end{enumerate}

We define an operation of substitution $[a/x]E$ that avoids variable capture.  Here $a$ is an object and $x$ a variable of the same category, and $E$ is either an object or a proposition.  We omit the details of the definition.

We write $\phi \leftrightarrow \psi$ for $(\phi \supset \psi) \wedge (\psi \supset \phi)$.

\pagebreak

We also define an equality relation on every category.  For any category $A$ and objects $a$ and $b$ of $A$, we define the proposition $a =_A b$ as follows.
\begin{itemize}
 \item If $A$ is a basic category, then $a =_A b$ is the proposition $a = b$.
\item If $A$ is $B \times C$ and either $B$ or $C$ is ideal, then $a =_A b$ is the proposition
\[ \pi_1(a) =_B \pi_1(b) \wedge \pi_2(a) =_C \pi_2(b) \enspace . \]
\item If $A$ is $B \boldarrow C$, then $a =_A b$ is the proposition
\[ \forall x. a(x) =_C b(x) \]
where $x$ is associated with $B$.
\item If $A$ is $\Set{B}$, then $a =_A b$ is the proposition
\[ \forall x(x \in a \leftrightarrow x \in b) \]
where $x$ is associated with $B$.
\end{itemize}

\subsubsection{Axioms}
The theorems of Weyl's system are those that can be derived via \emph{classical} predicate logic from the following axioms:
\begin{longenum}
\item For any basic category $A$,
\begin{gather*}
 \forall x. x = x \\
\forall x \forall y (x = y \supset \phi \supset [y/x] \phi)
\end{gather*}
where $\phi$ is any proposition, and $x$ and $y$ are associated with $A$.
\item
Peano's axioms for the natural numbers:
\begin{gather*}
\forall x \neg (\s(x) = 0) \\
\forall x \forall y (\s(x) = \s(y) \supset x = y) \\
{[0/x]}\phi \supset \forall x (\phi \supset [\s(x) / x] \phi) \supset \forall x \phi
\end{gather*}
for any proposition $\phi$, arithmetic or large.
\item
For any categories $A$ and $B$,
\begin{gather*}
 \forall x \forall y. \pi_1((x,y)) =_A x \\
\forall x \forall y. \pi_2((x,y)) =_B y
\end{gather*}
where $x$ is associated with $A$ and $y$ with $B$.
\item
For any categories $A$ and $B$, and any object $b$ of category $B$,
\[ \forall y((\lambda x.b)(y) =_B [y/x]b) \]
where $x$ and $y$ are associated with $A$.
\item
For any category $A$ and arithmetic proposition $\phi$,
\[ \forall y(y \in \{ x \mid \phi \} \leftrightarrow [y/x]\phi) \]
where $x$ and $y$ are associated with $A$.
\item For any category $A$ and object $f$ of category $A \boldarrow A$,
\begin{gather*}
 \forall x (\overline{f}(x,0) =_A x) \\
\forall x \forall n (\overline{f}((x,\s(n))) =_A \overline{f}((f(x),n)))
\end{gather*}
\end{longenum}

\paragraph{Definition by Recursion}
The operation of iteration allows us to define functions by primitive recursion.  Let $f : A \boldarrow A$ and $g : A \times A \times \mathbb{N} \boldarrow A$.  Suppose we want to define the function $h : A \times \mathbb{N} \boldarrow A$ such that
\begin{eqnarray*}
 h(a,0) & = & f(a) \\
h(a,n+1) & = & g(h(a,n),a,n)
\end{eqnarray*}
This can be done as follows.  Define $k : A \times A \times \mathbb{N} \boldarrow A \times A \times \mathbb{N}$ by
\[ k(x,y,n) = (g(x,y,n),y,n+1) \enspace . \]
Then
\[ h(a,n) = \pi_1^3(\overline{k}((f(a),a,0),n)) \enspace , \]
where $\pi_1^3(a,b,c) = a$.
%

\subsubsection{Extensions to Weyl's System}
\label{section:extension}
There are two features in our system which were not
explicitly present in Weyl's system, but which can justifiably be
seen as conservative extensions of the same.

Weyl did not have the categories $A \times B$ of pairs, and did not have all of the function categories $A \boldarrow B$ of our system.  Instead, apart from the basic categories, the categories of Weyl's system were those of the form
\[ \Set{B_1 \times \cdots \times B_n} \mbox{ and } A_1 \times \cdots \times A_m \boldarrow \Set{B_1 \times \cdots \times B_n} \]
in our notation.

Instead of a category $\mathbb{N} \boldarrow \mathbb{N}$, functions from $\mathbb{N}$ to $\mathbb{N}$ in \emph{Das Kontinuum} are sets of ordered pairs, of category $\Set{\mathbb{N} \times \mathbb{N}}$.

Weyl allowed the iteration only of functions from a category $\Set{A_1 \times \cdots \times A_n}$ to itself (the `Principle of Iteration' \cite[p.~36]{weyl:continuum}).  He showed by example how definition by recursion is then possible: addition is defined by iterating a suitable function from $\Set{\mathbb{N} \times \mathbb{N}}$ to $\Set{\mathbb{N} \times \mathbb{N}}$ \cite[p.~51]{weyl:continuum}, and multiplication is defined in an `entirely analogous' manner \cite[p.~53]{weyl:continuum}.

We have also deviated from Weyl's system in two more minor ways.  We have used a primitive operation $\s(x)$ for successor, where Weyl used a binary relation $Sxy$.
We choose to start the natural numbers at 0, where Weyl begins at 1.

\section{Weyl's Foundation as a Logic-Enriched Type Theory}
\label{section:lett}

A modern eye reading \emph{Das Kontinuum}
is immediately struck by how similar the system presented there is to what we now know as
a type theory; almost the only change needed is to replace the word
`category' with `type'.  In particular, Weyl's system is very
similar to a \emph{logic-enriched type theory} (LTT).

The LTT we shall construct, which we shall call \LTTW, must involve:
\begin{itemize}
 \item
 natural numbers, ordered pairs and functions;
\item  predicate logic;
\item the ability to prove propositions by induction;
\item the formation of \emph{sets}.
\end{itemize}
We shall need to divide our types into \emph{small} and \emph{large} types, corresponding to Weyl's basic and ideal categories.  We shall also need to divide our propositions into \emph{arithmetic} and \emph{large} propositions.  To make these divisions, we shall use a \emph{type universe} and a \emph{propositional universe}.

There exist today many \emph{logical frameworks}, designed as
systems for representing many different type theories.  It requires
only a small change to make a logical framework capable of
representing LTTs as well.
We have constructed $\LTTW$ within a variant of the logical framework
LF \cite{luo:car},
which is implemented by the proof checker Plastic \cite{plastic}.

\subsection{Logic-Enriched Type Theories in Logical Frameworks}

Recall that a logical framework, such as LF, is intended as a metalanguage for constructing various
type theories, the \emph{object systems}.  The frameworks consist of \emph{kinds} and \emph{objects}.  An object system is constructed in a framework by making certain \emph{declarations}, which extend the framework with new constants and rules of deduction.

A type theory divides mathematical objects, or \emph{terms}, into \emph{types}.  The framework LF provides a kind $\Type$ and a kind constructor $\mathrm{El}$.  The intention is that LF be used for constructing a type theory, with the types represented by the objects of kind $\Type$, and the terms of the type $A$ by the objects of kind $\El{A}$.  Further details can be found in \cite{luo:car}.

A \emph{logic-enriched type theory} has two new syntactic categories: besides terms and types, there are \emph{propositions} and \emph{proofs}.
%
To make LF capable of representing LTTs, we add a kind $\Prop$ and a
kind constructor $Prf$.  We shall refer to this extended framework as \LF.
The rules of deduction for these new kinds $\Prop$ and
$\Prf{\cdots}$ are given in Figure~\ref{fig:rules}, along with the
rules those for $\Type$ and $El$, for comparison. The full syntax and
rules of deduction of \LF~are given in Appendix \ref{appendix:lf}.

We construct an LTT in \LF~by representing:
\begin{itemize}
\item
the \emph{types} by the objects of kind $\Type$;
\item
the \emph{terms} of type $A$ by the objects of kind $\El{A}$;
\item
the \emph{propositions} by the objects of kind $\Prop$;
\item
the \emph{proofs} of the proposition $\phi$ by the objects of kind $\Prf{\phi}$.
\end{itemize}

\begin{figure}[top]
\begin{longenum}
\item
Rules of Deduction for $\Type$ and $El$
\[ \begin{prooftree}
\Gamma \vald
\justifies
\Gamma \vdash \Type \kind
   \end{prooftree}
\qquad
\begin{prooftree}
\Gamma \vdash A : \Type
\justifies
\Gamma \vdash \El{A} \kind
\end{prooftree}
\qquad
\begin{prooftree}
\Gamma \vdash A = B : \Type
\justifies
\Gamma \vdash \El{A} = \El{B}
\end{prooftree} \]
\item
Rules of Deduction for $\Prop$ and $Prf$
\[ \begin{prooftree}
\Gamma \vald
\justifies
\Gamma \vdash \Prop \kind
   \end{prooftree}
\qquad
\begin{prooftree}
\Gamma \vdash P : \Prop
\justifies
\Gamma \vdash \Prf{P} \kind
\end{prooftree}
\qquad
\begin{prooftree}
\Gamma \vdash P = Q : \Prop
\justifies
\Gamma \vdash \Prf{P} = \Prf{Q}
\end{prooftree} \]
\end{longenum}
\caption{Kinds $Type$ and $Prop$ in \LF}
\label{fig:rules}
\end{figure}

\subsubsection{Example}
When constructing an LTT in \LF,
we can include conjunction by making the following declarations:
\begin{eqnarray*}
 \wedge & : & \Prop \rightarrow \Prop \rightarrow \Prop \\
\wedge I & : & (p,q : \Prop) \Prf{p} \rightarrow \Prf{q} \rightarrow \Prf{\wedge p q} \\
\wedge E1 & : & (p,q : \Prop) \Prf{\wedge p q} \rightarrow \Prf{p} \\
\wedge E2 & : & (p,q : \Prop) \Prf{\wedge p q} \rightarrow \Prf{q}
\end{eqnarray*}
This has the effect of extending the logical framework with the
constants $\wedge$, $\wedge I$, $\wedge E1$ and $\wedge E2$.
The first allows propositions of the form $\phi \wedge \psi$ to be formed,
and the last three are the introduction and elimination rules.

\subsection{Natural Numbers, Products, Functions and Predicate Logic}

We can now proceed to construct a logic-enriched type theory $\LTTW$ that
corresponds to the foundational system Weyl presents in \emph{Das
Kontinuum}.
In the body of this paper, we shall describe a few of the declarations that comprise the specification of $\LTTW$ in \LF.  The full list of declarations is given in Appendix \ref{appendix:lttw}.

Our starting point is an LTT that contains, in its datatype
world, a type $\mathbb{N}$ of natural numbers, as well as
non-dependent product and function types $A \times B$ and $A
\boldarrow B$; and, in its logical world, classical predicate
logic.  We present some of the declarations involved in its
specification in Figure \ref{fig:ltt}, namely those involving
natural numbers and implication.  The rules for natural numbers include the elimination rule, which allows the definition of functions by recursion over $\mathbb{N}$, and the rule for proof by induction.

\begin{figure}[top]
\textbf{Natural Numbers}
\begin{gather*}
\begin{array}{rcl}
\mathbb{N} & : & \Type \\
0 & : & \mathbb{N} \\
\s & : & \mathbb{N} \rightarrow \mathbb{N} \\
\EN & : & (C : \mathbb{N} \rightarrow \Type) C 0 \rightarrow ((x : \mathbb{N}) C x \rightarrow  C (\s x)) \rightarrow (n : \mathbb{N}) C n \\
\IN & : & (P : \mathbb{N} \rightarrow \Prop) \Prf {P 0} \rightarrow((x : \mathbb{N})
\Prf {P x} \rightarrow \Prf{P (\s x)}) \rightarrow \\
& & \qquad (n : \mathbb{N}) \Prf {P n}
\end{array} \\
\begin{array}{rcl}
\EN\, C\, a\, b\, 0 & = & a : \El{C 0} \\
\EN\, C\, a\, b\, (\s n) & = & b\, n\, (E_\mathbb{N}\, C\, a\, b\, n) : \El{C (\s n)} \\
\end{array}
\end{gather*}
\textbf{Implication}
\begin{eqnarray*}
\supset & : & \Prop \rightarrow \Prop \rightarrow \Prop \\
\supset \! \mathrm{I} & : & (P : \Prop) (Q : \Prop) (\Prf{P} \rightarrow \Prf{Q}) \rightarrow \Prf{P \supset Q} \\
\supset \! \mathrm{E} & : & (P : \Prop) (Q : \Prop) \Prf{P \supset Q} \rightarrow \Prf{P} \rightarrow \Prf{Q}
\end{eqnarray*}
\textbf{Peirce's Law}
\[ \mathrm{Peirce} : (P : \Prop) (Q : \Prop) ((\Prf{P} \rightarrow \Prf{Q}) \rightarrow \Prf{P}) \rightarrow \Prf{P} \]
\caption{Some of the constants involved in the declaration of $\LTTW$ in \LF} \label{fig:ltt}
\end{figure}

\subsection{Type Universes and Propositional Universes}

We have introduced our collection of types.  We now divide them into
the small and large types.

We use a familiar device to do this: a \emph{type universe}. A type universe $U$
is a type whose objects are names of types\footnote{Such a universe is called a universe \emph{\`a la Tarski}, as opposed to a universe \emph{\`a la Russell}, where the objects of the universe are themselves types.}.  The types that have a name in $U$ are the
\emph{small} types, and those that do not (such as $U$ itself) as the
\emph{large} types. We also introduce a constructor $T$.
For each name $a : U$, the type $T(a)$ is the type named by $a$.

For our system, we need a
universe $U$ that contains a name for $\mathbb{N}$, and a method for
constructing a name for $A \times B$ out of a name for $A$ and a
name for $B$.  We also need to introduce a relation of equality for every small type.  These are both done by the constants declared in Figure \ref{fig:univs}.

\begin{figure}[top]
\begin{enumerate}
 \item \textbf{The Type Universe}
\begin{eqnarray*}
 U & : & \Type \\
T & : & U \rightarrow \Type \\
\hat{\mathbb{N}} & : & U \\
\hat{\times} & : & U \rightarrow U \rightarrow U \\
T \hat{\mathbb{N}} & = & \mathbb{N} : \Type \\
T(a \hat{\times} b) & = & T a \times T b : \Type
\end{eqnarray*}
\item
\textbf{Propositional Equality}
\begin{eqnarray*}
 \simeq & : & (A : U) T A \rightarrow T A \rightarrow \Prop \\
\simeq I & : & (A : U) (a : T A) \Prf{a \simeq_A a} \\
\simeq E & : & (A : U) (P : T A \rightarrow \Prop) (a,b : T A) \\
& & \quad \Prf{a \simeq_A b} \rightarrow \Prf{P a} \rightarrow \Prf{P b}
\end{eqnarray*}
\item
\textbf{The Propositional Universe}
\begin{eqnarray*}
 \p & : & \Prop \\
V & : & \Prf{\p} \rightarrow \Prop \\
\hat{\bot} & : & \Prf{\p} \\
\hat{\supset} & : & \Prf{\p} \rightarrow \Prf{\p} \rightarrow \Prf{\p} \\
\hat{\forall} & : & (a : U) (Ta \rightarrow \Prf{\p}) \rightarrow \Prf{\p} \\
\hat{\simeq} & : & (a : U) Ta \rightarrow Ta \rightarrow \Prf{\p} \\
V(\hat{\bot}) & = & \bot : \Prop \\
V(p \hat{\supset} q) & = & V p \supset V q : \Prop \\
V(\hat{\forall} a \, p) & = & \forall  (T a) \, [x : T a] V(px) : \Prop \\
V(s \hat{\simeq}_a t) & = & s \simeq_a t : \Prop
\end{eqnarray*}
\end{enumerate}
\caption{A type universe and a propositional universe}
\label{fig:univs}
\end{figure}

We also need to divide our propositions into the arithmetic propositions
and the large propositions.  To do so, we use the notion
in the logical world which is analagous to a type universe: a \emph{propositional
universe}.

We wish to introduce the collection `$\p$' of names of the
\emph{arithmetic} propositions; that is, the propositions that only
involve quantification over small types.  We also introduce the constructor $V$; given $P : \p$, $V(P)$ shall be the small proposition named by $P$.

It is not immediately
obvious where the collection `$\p$' should live.
We choose to declare $\p : \Prop$, and use the objects of kind $\Prf{\p}$ for the names of the small propositions.

Now, it must be admitted that
$\p$ is not conceptually a proposition; it does not assert any
relation to hold between any mathematical objects.
We could have declared $\p : \Type$ instead.
It makes no
practical difference for this formalisation which choice is made.

We chose to declare $\p : \Prop$ as this provides a pleasing symmetry with
$U$ and $\Type$, and $\p$ seems to belong more to the logical world than the datatype world.
Until more foundational work on LTTs has been done, we accept this compromise: $\p$ is
a `proposition', each of whose `proofs' is a name of an arithmetic
proposition.\footnote{Other alternatives would be to introduce a new
top-kind to hold $\p$, or to make $\p$ itself a top-kind.  We do not discuss these here.}  We discuss this matter further in Section \ref{section:anomalies}.

The declarations associated with $\p$ are given in Figure \ref{fig:univs}(3). Note that the
propositional universe provides us with our first examples of
computation rules for propositions.

\subsection{The Predicative Notion of Set}
\label{section:set}

We now have all the machinery necessary to be able to introduce
\emph{typed sets}.  For any type $A$, we wish to introduce the type
$\Set{A}$ consisting of all the sets that can be formed, each of
whose members is an object of type $A$.  (Thus we do not have any
sets of mixed type.)  We take a set to be introduced by an
\emph{arithmetic predicate} over $A$; that is, an object of kind $A \rightarrow
\Prf{\p}$, a function which takes objects of $A$ and returns (a name of) an
arithmetic proposition.

\pagebreak

We therefore include the following in our LTT:
\begin{longitem}
\item
Given any type $A$, we can form the type $\Set{A}$.  The terms of $\Set{A}$ are all the sets that can be formed whose elements are terms of type $A$.
\item
Given a \emph{arithmetic} proposition $\phi[x]$ with variable $x$ of type $A$, we can form the set $\{ x : A \mid \phi[x] \} : \Set{A}$.
\item
If $a : A$ and $X : \Set{A}$, we can form the proposition $a \in X$, which is arithmetic.
\item
Finally, we want to ensure that the elements of the set $\{ x : A \mid \phi[x] \}$ are precisely the terms $a$ such that $\phi[a]$ is true.  This is achieved by adding a \emph{computation rule on propositions}:
\begin{equation}
 \label{eq:membership}
a \in \{ x : A \mid \phi[x] \} \mbox{ computes to } \phi[a] \enspace .
\end{equation}
\end{longitem}
The type $\Set{A}$ is to be a large type; we do not provide any means for forming a name of $\Set{A}$ in $U$.

We therefore make the declarations in Figure \ref{fig:sets}.
In particular, the constants listed there allow us to form the following objects:
\begin{itemize}
 \item For every type $A$, the type $\Set{A}$.
\item For every type $A$ and name $p[x] : \Prf{\p}$ of an arithmetic proposition, the object $\set \, A \, ([x:A]p[x])$, which represents the set $\{ x : A \mid V(p[x]) \}$ in $\Set{A}$.
\item Given objects $a : A$ and $X : \Set{A}$, the object $\in A \, a \, X$, which represents a name of the proposition $a \in X$.
\end{itemize}
The computation rule (\ref{eq:membership}) is represented by declaring: $\in A \, a \, (\set \, A \, P)$ computes to $P\, a$.

\begin{figure}[top]
\begin{eqnarray*}
\mathrm{Set} & : & \Type \rightarrow \Type \\
\set & : & (A : \Type) (A \rightarrow \Prf{\p}) \rightarrow \Set{A} \\
\in & : & (A : \Type) A \rightarrow \Set{A} \rightarrow \Prf{\p} \\
\in \! A \, a \, (\set \, A \, P) & = & P a : \Prf{\p}
\end{eqnarray*}
\caption{The Predicative Notion of Set}
\label{fig:sets}
\end{figure}

%
%

\subsubsection{Remarks}

\begin{longenum}
 \item
We have introduced here a \emph{predicative} notion of set.  It is impossible in $\LTTW$ to perform a definition that is impredicative in Weyl's sense.
This restriction on which sets can be formed, proposed by Weyl and formalised in $\LTTW$, is by no means the only way of defining a predicative notion of set.
Historically, there have been a number of different ways of interpreting the concept of `predicativity', and a number of different predicative formal systems that all impose different restrictions on which sets and functions may be defined.  We discuss this matter further in Section \ref{section:predicativity}.
\item
It would be easy to modify the system to use instead the \emph{impredicative} notion of set.  This would involve changing the declaration of either $\p$, $U$ or $\mathrm{Set}$; the details are given in Section \ref{section:impredicative}.  When we declare both the predicative and the impredicative systems in this way, all the proofs in the predicative system can be reused without change in the impredicative.  Further discussion of all these points may be found in \cite{luo:LTT06}.
\end{longenum}

\subsection{Anomalies in $\LTTW$}
\label{section:anomalies}
The system $\LTTW$ has certain features that are anomalies, unwanted side-effects of our design choices.

We chose to make the type universe $U$ an object of kind $\Type$.  This means that the objects of kind $\Type$ no longer correspond to the categories of Weyl's system.  There are certain extra objects in $\Type$: $U$ itself, as well as $U \times U$, $U \boldarrow U$, $\mathbb{N} \boldarrow U$, and so forth.  We can therefore do more in $\LTTW$ than we can in Weyl's system, such as
define functions $\mathbb{N} \boldarrow U$ by recursion, and hence define meta-functions $\mathbb{N} \rightarrow\Type$ by recursion.  For example, we can define the meta-function $f : (\mathbb{N} \rightarrow \Type$ where
\[ f(n) = \overbrace{\mathbb{N} \times \mathbb{N} \times \cdots \times \mathbb{N}}^{n} \enspace . \]

We also chose to make $\p$ an object of kind $\Prop$.  This results in the objects of kind $\Prop$ no longer corresponding to the propositions of Weyl's system.  There are extra objects in $\Prop$: $\p$ itself, as well as $\p \supset \p$, $\forall x : \mathbb{N}. \p$, and so forth.  If we had placed $\p$ in $\Type$ instead, we would get more anomalous types, and the ability to define functions into $\p$ by recursion.

These anomalies would not arise if we made $U$ and $\p$ top-kinds, instead of placing them in $\Type$ and $\Prop$.
The choice of where to place the universes makes no practical difference for this formalisation, because we did not make use of any of these anomalous objects in our formalisation.  We have checked that our proof scripts still parse if we declare $\p : \Type$.

It seems likely that these anomalies are harmless.  We conjecture that $\LTTW$ is a conservative extension of the system in which $U$ and $\p$ are top-kinds.  However, it also seems likely that this would not be true for stronger LTTs.  What effect the placement of universes has in LTTs is a question that needs further investigation.

\section{Formalisation in Plastic}
\label{section:formal}

We have formalised this work in a version of the proof assistant
Plastic \cite{plastic}, modified by Paul Callaghan to be an
implementation of LF$'$.  We have produced a formalisation which
includes all the definitions and proofs of several of the results
from Weyl's book.

In Plastic, all lines that are to be parsed begin with the character
\texttt{>}; any line that does not is a comment line.  A constant
$c$ may be declared to have kind $(x_1 : K_1) \cdots (x_n : K_n) K$
by the input line
\[ \texttt{> [} c \texttt{[} x_1 \texttt{:} K_1 \texttt{]} \cdots \texttt{[} x_n \texttt{:} K_n \texttt{] : } K \texttt{];} \]
We can define the constant $c$ to be the object $[x_1 : K_1] \cdots
[x_n : K_n] k$ of kind $(x_1 : K_1) \cdots (x_n : K_n) K$ by writing
\[ \texttt{> [} c \texttt{[} x_1 \texttt{:} K_1 \texttt{]} \cdots \texttt{[} x_n \texttt{:} K_n \texttt{] = } k \texttt{ : } K \texttt{];} \]
In both these lines, the kind indicator $\mathtt{:} K$ is optional, and is usually omitted.

We can make any argument implicit by replacing it with a `meta-variable' \texttt{?},
indicating that we wish Plastic to infer its value.


\subsection{Results Proven}

\subsubsection{Peano's Fourth Axiom}
Peano's fourth axiom is the proposition
\[ \forall x : \mathbb{N}. \s(x) \neq 0 \enspace . \]
This can be proven in $\LTTW$ by taking advantage of the fact that we can define functions from $\mathbb{N}$ to $\Set{\mathbb{N}}$ by recursion.

Define the meta-function $f : \mathbb{N} \rightarrow \Set{\mathbb{N}}$ as follows.
\begin{eqnarray*}
f(0) & = & \emptyset \\
f(n+1) & = & \{ x : \mathbb{N} \mid \top \}
\end{eqnarray*}
Now, we have $0 \in f(n+1)$.  If $n+1 = 0$, we would have $0 \in f(0) = \emptyset$, which is a contradiction.  Therefore, $n+1 \neq 0$.

Peano's fourth axiom is often surprisingly difficult to prove in a type theory, requiring a universe or something equally strong.  This is true in logic-enriched type theories, too; if we remove $U$, $\p$ and $\mathrm{Set}$ from $\LTTW$, then Peano's fourth axiom cannot be proven in the resulting system.  (Proof: the resulting system has a model in which every type has exactly one object.)

\subsubsection{Cardinality of Sets}
\label{section:cardinality}

In Weyl's system, we can define the predicate `the set $X$ has
exactly $n$ members' in the following manner.

Given a basic category $A$, define the function $K : \mathbb{N}
\boldarrow \Set{\Set{A}}$ by recursion as follows.  The intention
is that $K(n)$ is the set of all sets $X : \Set{A}$ that have at
least $n$ members.
\begin{eqnarray*}
K(0) & = & \{ X : \Set{A} \mid \top \} \\
K(n+1) & = & \{ X : \Set{A} \mid \exists a:A (a \in X \wedge X \setminus \{a\} \in K(n)) \}
\end{eqnarray*}
In Plastic, this is done as follows:
\begin{quote}
 \begin{verbatim}
> [at_least_set [tau : U] = E_Nat ([_ : Nat] Set (Set (T tau)))
>     (full (Set (T tau)))
>     [n : Nat] [Kn : Set (Set (T tau))] set (Set (T tau))
>        [X : Set (T tau)] ex tau [a : T tau]
>           and (in (T tau) a X) (in ? (setminus' tau X a) Kn)];
\end{verbatim}
\end{quote}

We define the proposition `$X$ has at least $n$ members' to be $X \in K(n)$.
\begin{quote}
\begin{verbatim}
> [At_Least [tau : U] [X : Set (T tau)] [n : Nat]
>    = In ? X (at_least_set tau n)];
\end{verbatim}
\end{quote}

For $n$ a natural number, define the \emph{cardinal number} $\overline{n}$ to be $\{ x : \mathbb{N} \mid x < n \}$.
\begin{quote}
\begin{verbatim}
> [card [n : Nat] = set Nat [x : Nat] lt x n];
\end{verbatim}
\end{quote}
Define the \emph{cardinality} of a set $A$ to be $|A| = \{ n : \mathbb{N} \mid A \mbox{ has at least } \s(n) \mbox{ members} \}$.
\begin{quote}
\begin{verbatim}
> [cardinality [tau : U] [A : Set (T tau)]
>    = set Nat [n : Nat] at_least tau A (succ n)];
\end{verbatim}
\end{quote}
We can prove the following result:
\begin{quote}
The cardinality $|X|$ of a set $X$ is either $\{ x : \mathbb{N} \mid \top \}$ or $\overline{n}$ for some $n$.
\end{quote}
We thus have two classes of cardinal numbers: $\overline{n}$, for measuring the size of finite sets, and $\{ x : \mathbb{N} \mid \top \}$, which we denote by $\infty$, for measuring the size of infinite sets.  (There is thus only one infinite cardinality in \emph{Das Kontinuum}.)  We define `$X$ has exactly $n$ members' to be $|X| =_{\Set{\mathbb{N}}} \overline{n}$.
\begin{quote}
\begin{verbatim}
> [infty = full Nat];
> [Exactly [tau : U] [A : Set (T tau)] [n : Nat]
>    = Seteq Nat (cardinality tau A) (card n)];
\end{verbatim}
\end{quote}
\pagebreak
With these definitions, we can prove results such as the following:
\begin{enumerate}
\item
If $A$ has at least $n$ elements and $m \leq n$, then $A$ has at least $m$ elements.
\item
If $A$ has exactly $n$ elements, then $m \leq n$ iff $A$ has at least $m$ elements.
\item
If $A$ has exactly $m$ elements, $B$ has exactly $n$ elements, and $A$ and $B$ are disjoint, then $A \cup B$ has exactly $m + n$ elements.
\end{enumerate}

We have thus provided definitions of the concepts `having at least $n$ members' and `having exactly $n$ members' in such a way that the sets
$\{ X : \Set{A} \mid X \mbox{ has at least } n \mbox{ members} \}$ and $\{ X : \Set{A} \mid X \mbox{ has exactly } n \mbox{ members} \}$
are definable predicatively.  This would not be possible if, for example, we defined `$X$ has exactly $n$ elements' as the existence
of a bijection between $X$ and $\overline{n}$; we would have to quantify over the ideal category $A \boldarrow \mathbb{N}$.  It also cannot be
done as directly in a predicative system of second order arithmetic
such as ACA$_0$ \cite{simpson:sosoa}.

\subsubsection{Construction of the Reals}

The set of real numbers is constructed by the following process. We
first define the type of integers $\mathbb{Z}$, with a defined relation of equality $\approx_\mathbb{Z}$.
We then define a
\emph{rational} to be a pair of integers, the second of which is non-zero.  That is, for $q : \mathbb{Z} \times \mathbb{Z}$, we define `$q$ is rational' by
\[ \langle x,y \rangle \mbox{ is rational} \equiv y \not\approx_\mathbb{Z} 0 \enspace . \]
We proceed to define equality of rationals ($q \approx_\mathbb{Q} q'$), addition, multiplication and ordering on the rationals.

A \emph{real} is a Dedekind cut of rationals; that is, an object $R$
of the category $\Set{\mathbb{Z} \times \mathbb{Z}}$ that:
\begin{itemize}
\item
is a \emph{domain of rationals}; if $q$ and $q'$ are rationals, $q
\in R$, and $q \approx_\mathbb{Q} q'$, then $q' \in R$;
\item
is \emph{closed downwards}; if $q$ and $q'$ are rationals, $q \in R$, and $q' < q$, then $q' \in R$;
\item
has no maximal element; for every rational $q \in R$, there exists a rational $q' \in R$ such that $q < q'$;
\item
and is neither empty nor full; there exists a rational $q$ such that $q \in R$, and a rational $q'$ such that $q' \notin R$.
\end{itemize}
Equality of reals is defined to be extensional equality restricted to the rationals:
\[ R \approx_\mathbb{R} S \equiv \forall q (q \mbox{ is rational} \supset (q \in R \leftrightarrow q \in S)) \]

We note that, in this formalisation, we could define the collection of integers as a type, because every pair of natural numbers is an integer.  In contrast,
there was no way to define the collection of rationals as a type, say as the `sigma-type' `$(\Sigma q : \mathbb{Z} \times \mathbb{Z}) q \mbox{ is rational}$'.  This is because our LTT offers no way to form a type from a type $\mathbb{Z} \times \mathbb{Z}$ and a proposition `$q$ is rational'.  We are, however,  able to form the \emph{set} $\mathbb{Q} = \{ q : \mathbb{Z} \times \mathbb{Z} \mid q \mbox{ is rational} \} : \Set{\mathbb{Z} \times \mathbb{Z}}$.  Likewise, we cannot form the type of reals, but we can form the set of reals $\mathbb{R} : \Set{\Set{\mathbb{Z} \times \mathbb{Z}}}$.

\subsubsection{Real Analysis}

Weyl was keen to show that his predicative system was strong enough to be used for mathematical work by demonstrating that, while several traditional theorems cannot be proven within it, we can usually prove a version of the theorem that is only slightly weaker.

For example, we cannot predicatively prove the \emph{least upper bound principle}: that every set $A$ of real numbers bounded above has a least upper bound $l$.
 Impredicatively, we would define $l$ to be the union of $A$.
This cannot be done predicatively, as it involves quantification over real numbers.  However, we can prove the following two statements, which are enough for most practical purposes:
\begin{enumerate}
\item
Every set $S$ of \emph{rational} numbers bounded above has a unique (real) least upper bound $l$.  Take
$l = \{ q \in \mathbb{Q} \mid (\exists q' \in S) q < q' \}$.
\item
Every \emph{sequence} $r_1, r_2, \ldots$ of real numbers bounded above has a unique least upper bound $l$.  Take
$l = \{ q \in \mathbb{Q} \mid (\exists n : \mathbb{N}) q \in r_n \}$.
\end{enumerate}
These involve only quantification over the rationals and the natural numbers, respectively.  (We note that either of these is equivalent to the least upper bound principle in an impredicative setting.)

The first is enough to prove the classical Intermediate Value Theorem:
\begin{quote}
If $f : \Set{\mathbb{Z} \times \mathbb{Z}} \boldarrow \Set{\mathbb{Z} \times \mathbb{Z}}$ is a continuous function from the reals to the reals, and
$f(a) < v < f(b)$
for some reals $a$, $b$, $v$ with $a < b$, then there exists a real $c$ such that $a < c < b$ and $f(c) = v$.
\end{quote}
The predicative proof of this theorem takes advantage of the fact that a continuous function is determined by its values on the rationals.
Weyl defined $c$ to be the least upper bound of the set of all \emph{rationals} $q$ such that $a < q < b$ and $f(q) < v$.  In the formalisation, we defined directly: $c = \{ q \in \mathbb{Q} \mid (\exists q' \in \mathbb{Q}) (q < q' < b \wedge f(q') < v) \}$.

\subsection{An Impredicative Development}
\label{section:impredicative}

As mentioned in Section \ref{section:set}, it is not difficult
to modify this formulation to get a development of the same theorems
in an impredicative system.  All we have to do is remove the
distinction between large and arithmetic propositions.

We do this by adding two impredicative quantifiers to $\p$, together with their computation rules:
\begin{xalignat*}{2}
\overline{\forall} & : (A : \Type) (A \rightarrow \Prf{\p}) \rightarrow \Prf{\p} & V(\overline{\forall} A P) & = \forall A ([x : A] V (Px)) \\
\overline{\exists} & : (A : \Type) (A \rightarrow \Prf{\p}) \rightarrow \Prf{\p} & V(\overline{\exists} A P) & = \exists A ([x : A] V (Px))
\end{xalignat*}

Now $\p$, which determines the collection of propositions over which
sets can be formed, covers the whole of $\Prop$.  We can form the set
$\{ x : A \mid \phi[x] \}$ for any well-formed proposition $\phi[x]$.
However, all our old proof files written in terms of $\p$ and $V$
still parse.

Once this change has been made, we can go on to prove the statement
that every set of real numbers bounded above has a least upper
bound.

\pagebreak
There are other ways in which we could have made our system impredicative.  We could have changed the construction of $U$,
so that every type in $\Type$ has a name.  The simplest way would be to erase the universes $U$ and $\p$, and to change every instance of $\p$ to $\Prop$ in the constants associated with $\mathrm{Set}$ (Figure \ref{fig:sets}).  However, this would leave us unable to reuse the proofs that we had written for the predicative system.
We could also have replaced the declarations of $\p$ and $V$ (Figure~\ref{fig:univs}) with
\begin{xalignat*}{2}
\p & = \Prop &
V & = [x : \Prop] \Prop \\
\hat{\bot} & = \bot &
\hat{\supset} & = \supset \\
\hat{\forall} & = [A : U] \forall (T A) &
\hat{\simeq} & = \simeq
\end{xalignat*}
However, at present Plastic becomes unstable when equations are declared at the top-kind level such as $\p = \Prop$.


\section{Related Work}
\label{section:predicativity}
Weyl's work in \emph{Das Kontinuum} is by no means the only attempt to provide a predicative foundation to mathematics.
Historically, there have been several different views on which definitions and proofs are predicatively acceptable.  Of these, Weyl's in \emph{Das Kontinuum} is one of the strictest.  Many of these have been formalised as systems of second-order arithmetic.  The question of to what extent they may be formalised by type theories, and the related question of which type theories may be considered predicative, have not received much attention before now.

\subsection{Other Formalisations of \emph{Das Kontinuum}}

Feferman \cite{feferman:kontinuum} has claimed that the system ACA$_0$ is ``a modern formulation of Weyl's system.''  The system ACA$_0$ is a subsystem of second-order arithmetic, so it deals only with natural numbers and sets of natural numbers.  A set $\{x \mid \phi[x]\}$ may only be introduced in ACA$_0$ for an \emph{arithmetic} predicate $\phi[x]$: this is the `arithmetic comprehension axiom' that gives the system its name.  The stronger system ACA has also been studied, which has a stronger induction principle: ACA$_0$ has the induction axiom
\[ \forall X (0 \in X \supset \forall x (x \in X \supset \s(x) \in X) \supset \forall x x \in X) \enspace , \]
whereas ACA has the induction \emph{schema}
\[ \phi[0] \supset \forall x (\phi[x] \supset \phi[\s(x)]) \supset \forall x \phi[x] \]
for every formula $\phi[x]$.

Weyl does go beyond ACA$_0$ in two places: in his definition of the cardinality of a set (\cite[Chapter 1 \S 7, pp. 38--39]{weyl:continuum}), and the results proven using this definition (\cite[Chapter 2 \S 1, pp.55f]{weyl:continuum}).  The definition requires the use of third-order sets (sets of sets), and the proofs use a stronger induction principle than the one found in ACA$_0$.  When formalising these parts of the book (see Section \ref{section:cardinality} above), we needed to use the type $\Set{\Set{A}}$, and to use $\IN$ with a large proposition.

\pagebreak

The rest of \emph{Das Kontinuum} can be formalised in ACA$_0$.  It is also possible to find another definition of the cardinality of a set that can be formalised in ACA$_0$, and the results in \emph{Das Kontinuum} can then be proven in ACA$_0$.  However, these two sections of the book show that Weyl did intend his system to be more than second-order.

\subsection{Other Approaches to Predicativity}


The concept of predicativity originated with Poincar\'e and Russell in 1906--8, who
held that the paradoxes of set theory each invoked a definition that involves a \emph{vicious circle}; in each case, a set $A$ is defined using a concept that presupposes the set $A$ itself.  They proposed the \emph{vicious circle principle}, which holds that such definitions are illegitimate.  Poincar\'e wrote \cite{poincare:ml3}: ``A definition containing a vicious circle defines nothing.''  Russell expressed the vicious circle principle as follows \cite{russell:pl}: ``Whatever involves an apparent [bound] variable must not be among the possible values of that variable.''

Russell proceeded to develop the \emph{theory of types} \cite{russell:mlbtt}, a system of logic that strictly adheres to the vicious circle principle.  The theory divides mathematical objects into types.  These include the type of individuals; the type of 0th-order or \emph{arithmetic} sets of individuals, sets that can be defined using quantifiers that range over individuals only; the type of 1st-order sets, sets that can be defined by quantifying over individuals and 0th-order sets; the type of 2nd-order sets, defined by quantification over individuals, 0th-order sets and 1st-order sets; and so forth\footnote{Russell's theory of types included many more types than those listed here.  For a historical account, we refer to \cite{kln:tlm1940}.}.

This ramification has some undesirable consequences; for example, if we define real numbers as sets of rationals, we find we have 0th-order reals, 1st-order reals, and so forth.  The usual development of analysis cannot be carried out in this framework: for example, the least upper bound property does not hold for any of these reals.

Russell's solution was to introduce the \emph{Axiom of Reducibility}: for every $n$th-order set, there is a 0th-order set with the same members.  This axiom effectively allows impredicative definitions to be made in the theory of types.  Weyl's solution, as we have seen, was to allow only 0th-order sets.

G\"odel \shortcite{godel:rml} suggested constructing sets of \emph{transfinite} order; to form sets of order $\alpha$, where $\alpha$ ranges over some initial segment of the ordinals.  The sets of order $\alpha + 1$ are those that can be formed by quantifying over the sets of order $\leq \alpha$; for each limit ordinal $\lambda$, the sets of order $\lambda$ are the union of the sets of lower order.  This idea was later named the \emph{ramified analytic hierarchy} by Kleene \shortcite{kleene:hntp}.  Kreisel \shortcite{kreisel:predicativite} that the predicatively acceptable sets should be identified with the sets of order $\alpha$ for $\alpha$ a recursive ordinal.  Feferman \shortcite{feferman:spa} and Sch\"utte \shortcite{schutte:pwo} independently suggested that the predicatively acceptable sets should be identified with those of order $< \Gamma_0$.  This latter idea has dominated the thinking on predicativity since.
\pagebreak

\subsection{Predicativity in Type Theory}

In general, the type theory community has confined its interest in predicativity to the question of whether a \emph{universe} is predicative.

Roughly, a universe is a type of types.  A universe $U$ is called \emph{impredicative} if some of the types in $U$ are constructed using $U$ itself; otherwise, $U$ is \emph{predicative}.  A type theory is called \emph{predicative} if it uses no impredicative universes.

Martin-L\"of's original type theory \cite{ML:71} used a single
universe $V$, which was highly impredicative as the system used the
typing rule $V \in V$.  After Girard \shortcite{girard:thesis}
showed that the system was inconsistent, Martin-L\"of \shortcite{ml:tt}
modified the system so that the universe became predicative. Later
he extended the system with an infinite sequence of universes
\shortcite{ml:ittpp}, and with \emph{W-types}, types of well-ordered
trees  \cite{ml:cmcp}.

Apart from the universes and the W-types, all the types in
Martin-L\"of type theory are simple inductive definitions, and so
can be defined by an explicit predicative definition given the
natural numbers; in this sense, Martin-L\"of type theory without
W-types is predicative modulo the natural numbers.  It was
conjectured by Peter Hancock \cite{ml:ittpp} and proved independently by Peter
Aczel and Feferman \cite{feferman:iifpt} that the proof-theoretic
strength of Martin-L\"of type theory, with infinitely many universes
but without W-types, is $\Gamma_0$.

Including the W-types  increases the strength of the system dramatically.  Setzer \shortcite{setzer:thesis} has shown that the theory with W-types and just one universe has a proof-theoretic ordinal much larger than $\Gamma_0$.  If one agrees with Feferman and Sch\"utte, then W-types are an impredicative element in the type theory.

It is worth noting that consistent impredicative type theories have also been usefully employed, most notably the Calculus of Constructions \cite{ch:coc} and its variants.

\subsection{Predicativity in Logic-Enriched Type Theories}

Logic-enriched type theories may provide a useful tool for investigating all these different conceptions of predicativity, as they provide a setting in which the three questions listed at the start of the section are kept separate.  When designing a logic-enriched type theory, we are able to adjust three parameters independently.
\begin{enumerate}
 \item We can make the propositional universes stronger or weaker.  This will determine which predicates may be used to define sets.
\item We can include more or fewer types in each type universe.  This will determine which inductive definitions may be used to define types.
\item We can make the universes predicative or impredicative.
\end{enumerate}
This flexibility means logic-enriched type theories form a very rich structure.  It would likely be fruitful to explore this structure by constructing LTTs corresponding to different approaches to the foundations of mathematics, and investigating correspondences between these LTTs and both traditional type theories and systems of predicate logic.  This would provide a setting in which we could experiment by (say) adding and removing generalised inductive types, in a way that is much more suitable for machine-assisted formalisation than are systems of predicate logic.

\section{Conclusion}

We have conducted a case study in Plastic of the use of a type-theoretic
framework to construct a logic-enriched type theory as the basis for
a formalisation of a non-constructive system of mathematical
foundations, namely that presented in Weyl's \textit{Das Kontinuum}.
As a representation of Weyl's work, it is arguably better in some
ways than such second-order systems as ACA$_0$
\cite{feferman:kontinuum}, since we can form a definition of the
cardinality of a set that is much closer to Weyl's own.  The
formalisation work required only a minor change to the existing
logical framework implemented in Plastic.

\subsection{Further Work}

We have seen how $\LTTW$ corresponds very closely to Weyl's system.  It is possible to embed the system ACA in $\LTTW$. We can also embed the system ACA$_0$ in $\LTTW$ with the constant $\IN$ removed and replaced with
\begin{eqnarray*}
\mathrm{ind}_\mathbb{N} & : & (P : \mathbb{N} \rightarrow \Prf{\p}) \Prf{V(P0)} \rightarrow  \\
& & \quad ((x:\mathbb{N})\Prf{V(Px)} \rightarrow \Prf{V(P(\s x))}) \rightarrow (n : \mathbb{N}) \Prf{V(Pn)} \enspace ,
\end{eqnarray*}
which allows us to prove only \emph{arithmetic} propositions by induction.

It is also possible to define a translation from $\LTTW$ to Martin-L\"of Type Theory with one universe extended by the axiom of excluded middle, and from ACA to $\LTTW$.
Future work will involve investigating these translations further --- whether they are conservative, and whether a translation can be defined in the opposite direction in each case.

We also hope to define more logic-enriched type theories corresponding to other schools in the foundations of mathematics; both other approaches to predicativity, and other schools entirely.  For example, we anticipate we can construct a system corresponding to the Theory of Types by extending $\LTTW$ with infinitely many type and propositional universes.  A stronger universe construction should also allow us to capture Kreisel's construction of the ramified analytic hierarchy over the recursive ordinals.  It remains to be seen whether we can capture Feferman-Sch\"utte predicativity, or whether another principle more naturally suited to LTTs will be discovered.

It would also be interesting to
carry out the impredicative development of analysis in our setting,
reusing the code from the predicative development.


\appendix
\section{The Logical Framework \LF}
\label{appendix:lf}

\LF\ is an extension of the logical framework LF as described in
Chapter 9 of \cite{luo:car}.  LF is a typed version of
Martin-L\"{o}f's logical framework and \LF\ extends LF by the
propositional kind $\Prop$ and the associated rules.

Formally, the terms of \LF\ are of the following forms:
\[ \Type,\ El(A),\ \Prop, \Prf{P},\ (x:K)K',\ [x:K]k',\ f(k), \]
where the free occurrences of variable $x$ in $K'$ and $k'$ are
bounded by the binding operators $(x:K)$ and $[x:K]$, respectively.  We shall usually omit $El$ when writing kinds.  We write $K \rightarrow K'$ for $(x:K)K'$ when $x$ does not occur free in $K'$.
\pagebreak

There are five forms of judgements:
\begin{itemize}
\item \ $\Gamma \vald$, which asserts that $\Gamma$ is a valid context;
\item
\ $\Gamma \vdash K \kind$, which asserts that $K$ is a kind;
\item
\ $\Gamma \vdash k\ \colon K$, which asserts that $k$ is an object
of kind $K$;
\item
\ $\Gamma \vdash k=k'\ \colon K$, which asserts that $k$ and $k'$
are equal objects of kind $K$;
\item
\ $\Gamma \vdash K=K'$, which asserts that $K$ and $K'$ are equal
kinds.
\end{itemize}
The rules of deduction of \LF\ are those of LF \cite{luo:car} plus
those involving $\Prop$.  For completeness, they are given in
Figures~\ref{LF-rulesGen} and~\ref{LF-rulesSpec}.

\begin{figure}
\begin{minipage}{\linewidth}
\ \\

\ \ \ \emph{Contexts and assumptions}
$$
\frac{}{\langle\rangle\: \: valid}\: \: \: \frac{\Gamma \vdash K\:
\: kind\: \: \: x\notin FV(\Gamma )}{\Gamma ,x:K\: \: valid}\: \: \:
\frac{\Gamma ,x:K,\Gamma '\: \: valid}{\Gamma ,x:K,\Gamma '\vdash
x\colon K}$$

\ \ \ \emph{General equality rules}
$$
\frac{\Gamma \vdash K\: \: kind}{\Gamma \vdash K=K}\: \: \:
\frac{\Gamma \vdash K=K'}{\Gamma \vdash K'=K}\: \: \: \frac{\Gamma
\vdash K=K'\: \: \Gamma \vdash K'=K''}{\Gamma \vdash K=K''}$$
$$
\frac{\Gamma \vdash k\colon K}{\Gamma \vdash k=k\colon K}\: \: \:
\frac{\Gamma \vdash k=k'\colon K}{\Gamma \vdash k'=k\colon K}\: \:
\: \frac{\Gamma \vdash k=k'\colon K\: \: \Gamma \vdash k'=k''\colon
K}{\Gamma \vdash k=k''\colon K}$$

\ \ \ \emph{Equality typing rules}
$$
\frac{\Gamma \vdash k\colon K\: \: \Gamma \vdash K=K'}{\Gamma \vdash
k\colon K'}\: \: \: \frac{\Gamma \vdash k=k'\colon K\: \: \Gamma
\vdash K=K'}{\Gamma \vdash k=k'\colon K'}$$

\ \ \ \emph{Substitution rules}
$$
\frac{\Gamma ,x:K,\Gamma '\: \: valid\: \: \Gamma \vdash k\colon
K}{\Gamma ,[k/x]\Gamma '\: \: valid}$$
$$
\frac{\Gamma ,x:K,\Gamma '\vdash K'\: \: kind\: \: \Gamma \vdash
k\colon K}{\Gamma ,[k/x]\Gamma '\vdash [k/x]K'\: \: kind}\: \: \:
\frac{\Gamma ,x:K,\Gamma \vdash K'\: \: kind\: \: \Gamma \vdash
k=k'\colon K}{\Gamma ,[k/x]\Gamma '\vdash [k/x]K'=[k'/x]K'}$$
$$
\frac{\Gamma ,x:K,\Gamma '\vdash k'\colon K'\: \: \Gamma \vdash
k\colon K}{\Gamma ,[k/x]\Gamma '\vdash [k/x]k'\colon [k/x]K'}\: \:
\: \frac{\Gamma ,x:K,\Gamma '\vdash k'\colon K'\: \: \Gamma \vdash
k_{1}=k_{2}\colon K}{\Gamma ,[k_{1}/x]\Gamma '\vdash
[k_{1}/x]k'=[k_{2}/x]k'\colon [k_{1}/x]K'}$$
$$
\frac{\Gamma ,x:K,\Gamma '\vdash K'=K''\: \: \Gamma \vdash k\colon
K}{\Gamma ,[k/x]\Gamma '\vdash [k/x]K'=[k/x]K''}\: \: \:
\frac{\Gamma ,x:K,\Gamma '\vdash k'=k''\colon K'\: \: \Gamma \vdash
k\colon K}{\Gamma ,[k/x]\Gamma '\vdash [k/x]k'=[k/x]k''\colon
[k/x]K'}$$

\
\end{minipage}
\caption{Rules of deduction of \LF\ (I)} \label{LF-rulesGen}
\end{figure}

\begin{figure}
\begin{minipage}{\linewidth}
\ \\

\ \ \ \emph{The kind $\Type$}
$$
\frac{\Gamma \: \: valid}{\Gamma \vdash \Type\: \: kind}\: \: \:
\frac{\Gamma \vdash A\colon \Type}{\Gamma \vdash El(A)\: \: kind}\:
\: \: \frac{\Gamma \vdash A=B\colon \Type}{\Gamma \vdash
El(A)=El(B)}$$

\ \ \ \emph{The kind $\Prop$}
$$
\frac{\Gamma \: \: valid}{\Gamma \vdash \Prop\: \: kind}\: \: \:
\frac{\Gamma \vdash P\colon \Prop}{\Gamma \vdash \Prf{P}\: \:
kind}\: \: \: \frac{\Gamma \vdash P=Q\colon \Prop}{\Gamma \vdash
\Prf{P}=\Prf{Q}}$$

\ \ \ \emph{Dependent product kinds}
$$
\frac{\Gamma \vdash K\: \: kind\: \: \Gamma ,x:K\vdash K'\: \:
kind}{\Gamma \vdash (x:K)K'\: \: kind}\: \: \: \frac{\Gamma \vdash
K_{1}=K_{2}\: \: \Gamma ,x:K_{1}\vdash K_{1}'=K_{2}'}{\Gamma \vdash
(x:K_{1})K_{1}'=(x:K_{2})K_{2}'}$$
$$
\frac{\Gamma ,x:K\vdash k\colon K'}{\Gamma \vdash [x:K]k\colon
(x:K)K'}\: \: \: \: \: \: \frac{\Gamma \vdash K_{1}=K_{2}\: \: \:
\Gamma ,x:K_{1}\vdash k_{1}=k_{2}\colon K}{\Gamma \vdash
[x:K_{1}]k_{1}=[x:K_{2}]k_{2}\colon (x:K_{1})K}$$
$$
\frac{\Gamma \vdash f\colon (x:K)K'\: \: \Gamma \vdash k\colon
K}{\Gamma \vdash f(k)\colon [k/x]K'}\: \: \: \frac{\Gamma \vdash
f=f'\colon (x:K)K'\: \: \Gamma \vdash k_{1}=k_{2}\colon K}{\Gamma
\vdash f(k_{1})=f'(k_{2})\colon [k_{1}/x]K'}$$
$$
\: \frac{\Gamma ,x:K\vdash k'\colon K'\: \: \Gamma \vdash k\colon
K}{\Gamma \vdash ([x:K]k')(k)=[k/x]k'\colon [k/x]K'}\: \: \: \:
\frac{\Gamma \vdash f\colon (x:K)K'\: \: x\notin FV(f)}{\Gamma
\vdash [x:K]f(x)=f\colon (x:K)K'}$$
\\
\end{minipage}
\caption{Rules of deduction of \LF\ (II)} \label{LF-rulesSpec}
\end{figure}

\section{Specification of $\LTTW$}
\label{appendix:lttw}

The following is the list of all the constant and equation
declarations that comprise the specification of $\LTTW$ within \LF.
We write the constants $\supset$, $\hat{\supset}$, $\times$ and
$\hat{\times}$ as infix; so we write $\supset \, \phi \, \psi$ as
$\phi \supset \psi$.  We shall also write $\simeq \, A \, a \, b$ as
$a \simeq_A b$, and $\hat{\simeq} \, A \, a \, b$ as $a
\hat{\simeq}_A b$.

\subsection{Classical Predicate Logic}
\begin{eqnarray*}
\bot & : & \Prop \\
\bot E & : & (p : \Prop) \Prf{\bot} \rightarrow \Prf{p} \\
\end{eqnarray*}
\begin{eqnarray*}
\supset & : & \Prop \rightarrow \Prop \rightarrow \Prop \\
\supset I & : & (p,q : \Prop) (\Prf{p} \rightarrow \Prf{q}) \rightarrow \Prf{p \supset q} \\
\supset E & : & (p,q : \Prop) \Prf{p \supset q} \rightarrow \Prf{p} \rightarrow \Prf{q} \\
\end{eqnarray*}
\begin{eqnarray*}
\mathrm{Peirce} & : & (p,q : \Prop) ((\Prf{p} \rightarrow \Prf{q}) \rightarrow \Prf{p}) \rightarrow \Prf{p} \\
\end{eqnarray*}
\begin{eqnarray*}
\forall & : & (A : \Type) (A \rightarrow \Prop) \rightarrow \Prop \\
\forall I & : & (A : \Type) (P : A \rightarrow \Prop) ((x : A) \Prf{Px}) \rightarrow \Prf{\forall A P} \\
\forall E & : & (A : \Type) (P : A \rightarrow \Prop) (a : A) \Prf{\forall A P} \rightarrow \Prf{Pa} \\
\end{eqnarray*}
\subsection{Natural Numbers}
\begin{eqnarray*}
\mathbb{N} & : & \Type \\
0 & : & \mathbb{N} \\
\s & : & \mathbb{N} \rightarrow \mathbb{N} \\
E_\mathbb{N} & : & (C : \mathbb{N} \rightarrow \Type) C 0 \rightarrow ((n : \mathbb{N}) C n \rightarrow C (\s n)) \rightarrow (n : \mathbb{N}) C n \\
\lefteqn{E_\mathbb{N} \, C \, a \, b \, 0 = a : C 0} \\
\lefteqn{E_\mathbb{N} \, C \, a \, b \, (\s n)
  =  b \, n \, (E_\mathbb{N} \, C \, a \, b \, n) : C (\s n)} \\
\IN & : & (P : \mathbb{N} \rightarrow \Prop) \Prf{P 0} \rightarrow ((n : \mathbb{N}) \Prf{Pn} \rightarrow \Prf{P(\s n)}) \rightarrow \\
& & \quad (n : \mathbb{N}) \Prf{Pn} \\
\end{eqnarray*}
\subsection{Pairs}
\begin{eqnarray*}
\times & : & \Type \rightarrow \Type \rightarrow \Type \\
\mathrm{pair} & : & (A, B  : \Type) A \rightarrow B \rightarrow A \times B \\
E_\times & : & (A,B : \Type) (C : A \times B \rightarrow \Type) ((a : A) (b : B) C (\mathrm{pair} \, A \, B \, a \, b)) \rightarrow \\
& & \quad (p : A \times B) C p \\
\lefteqn{E_\times \, A \, B \, C \, e \, (\mathrm{pair} \, A \, B \, a \, b) = e a b : C (\mathrm{pair} \, A \, B \, a \, b)} \\
\IX & : & (A,B : \Type) (P : A \times B \rightarrow \Type) ((a : A) (b : B) \Prf{P(\mathrm{pair} \, A \, B \, a \, b)}) \\
& & \quad \rightarrow (p : A \times B) \Prf{Pp} \\
\end{eqnarray*}
\subsection{Functions}
\begin{eqnarray*}
\boldarrow & : & \Type \rightarrow \Type \rightarrow \Type \\
\lambda & : & (A,B : \Type) (A \rightarrow B) \rightarrow (A \boldarrow B) \\
E_{\boldarrow} & : & (A, B : \Type) (C : (A \boldarrow B) \rightarrow \Type) ((b : A \rightarrow B) C (\lambda \, A \, B \, b)) \rightarrow \\
& & \quad (f : A \boldarrow B) C f \\
\lefteqn{E_{\boldarrow} \, A \, B \, C \, e \, (\lambda \, A \, B \, b) = e b : C (\lambda \, A \, B \, b)} \\
\IA & : & (A, B : \Type) (P : (A \boldarrow B) \rightarrow \Prop) ((b : A \rightarrow B) \Prf{P (\lambda \, A \, B \, b)}) \rightarrow \\
& & \quad (f : A \boldarrow B) \Prf{P f}
\end{eqnarray*}
\subsection{The Type Universe}
\begin{eqnarray*}
U & : & \Type \\
T & : & U \rightarrow \Type \\
\hat{\mathbb{N}} & : & U \\
\hat{\times} & : & U \rightarrow U \rightarrow U \\
\lefteqn{T \hat{\mathbb{N}} = \mathbb{N} : \Type} \\
\lefteqn{T (A \hat{\times} B) = T A \times T B : \Type}
\end{eqnarray*}
\subsection{Equality}
\begin{eqnarray*}
\simeq & : & (A : U) T A \rightarrow T A \rightarrow \Prop \\
\simeq I & : & (A : U) (a : T A) \Prf{a \simeq_A a} \\
\simeq E & : & (A : U) (P : T A \rightarrow \Prop) (a, b : T A) \Prf{P a} \rightarrow \Prf{a \simeq_A b} \rightarrow \Prf{P b}
\end{eqnarray*}
\subsection{The Propositional Universe}
\begin{eqnarray*}
\p & : & \Prop \\
V & : & \Prf{\p} \rightarrow \Prop \\
\hat{\bot} & : & \Prf{\p} \\
\hat{\supset} & : & \Prf{\p} \rightarrow \Prf{\p} \rightarrow \Prf{\p} \\
\hat{\forall} & : & (A : U) (T A \rightarrow \Prf{\p}) \rightarrow \Prf{\p} \\
\hat{\simeq} & : & (A : U) T A \rightarrow T A \rightarrow \Prf{\p} \\
\lefteqn{V \hat{\bot} = \bot : \Prop} \\
\lefteqn{V (p \hat{\supset} q) = V p \supset V q : \Prop} \\
\lefteqn{V (\hat{\forall} \, A \, P) = \forall (T A) [x : T A] V (P x) : \Prop} \\
\lefteqn{V (a \hat{\simeq}_A b) = a \simeq_A b : \Prop}
\end{eqnarray*}
\subsection{The Predicative Notion of Set}
\begin{eqnarray*}
\mathrm{Set} & : & \Type \rightarrow \Type \\
\set & : & (A : \Type) (A \rightarrow \Prf{\p}) \rightarrow \Set{A} \\
\in & : & (A : \Type) A \rightarrow \Set{A} \rightarrow \Prf{\p} \\
\lefteqn{\in \, A \, a \, (\set \, A \, P) = P a : \Prf{\p}}
\end{eqnarray*}

\begin{acks}
Thanks go to Paul Callaghan for his efforts in extending Plastic to
implement the type-theoretic framework which has made the reported
case study possible, and Peter Aczel for his comments during his
visit to Royal Holloway. Thanks also go to Thierry Coquand and the
anonymous referees, of this paper and the conference version, for
their very helpful comments.
\end{acks}

\bibliography{type}
\begin{received}
Received September 2008;
accepted December 2008
\end{received}

\end{document}